\documentclass[12pt]{article}
\usepackage{amsmath}
\usepackage{amssymb}
\usepackage{amsfonts}

\newcommand{\field}[1]{\mathbb{#1}}
\newcommand{\Z}{\field{Z}}
\newcommand{\C}{\field{C}}

\title{MINIMAL BOSONIZATION OF DOUBLE-GRADED QUANTUM MECHANICS}
\author{C.\ QUESNE\thanks{E-mail address: Christiane.Quesne@ulb.be} \\
{\small\sl Physique Nucl\'eaire Th\'eorique et Physique Math\'ematique, 
Universit\'e Libre de Bruxelles,} \\ 
{\small \sl Campus de la Plaine CP229, Boulevard~du Triomphe, B-1050
Brussels, Belgium}}
\date{ }
\begin{document}
\baselineskip=22pt plus 1pt minus 1pt
\maketitle

\begin{abstract}
The superalgebra of $\Z_2^2$-graded supersymmetric quantum mechanics is shown to be realizable in terms of a single bosonic degree of freedom. Such an approach is directly inspired by a description of the corresponding $\Z_2$-graded superalgebra in the framework of a Calogero-Vasiliev algebra or, more generally, of a generalized deformed oscillator algebra. In the case of the $\Z_2^2$-graded superalgebra, the central element $Z$ has the property of distinguishing between degenerate eigenstates of the Hamiltonian. 
\end{abstract}

\vspace{0.5cm}


 
\newpage
%
%
\section{Introduction}

In recent years, there has been a renewed interest in the application of a color superalgebra of $\Z_2^2 \equiv \Z_2 \otimes \Z_2$ grading  (and more generally of $\Z_2^n$ grading) to quantum mechanics. Such superalgebras were introduced in \cite{rittenberg78a, rittenberg78b} and extended the notion of ordinary ($\Z_2$-graded) Lie superalgebras appearing in \cite{kac, scheunert} (for an introduction to color superalgebras see, e.g., \cite{aizawa18}). Up to now, two independent, but related, lines of research have been followed: the study of physical models possessing a $\Z_2^2$-graded symmetry (see, e.g., \cite{aizawa16, bruce20a, aizawa20a, aizawa21, bruce20b}) and the investigation of parastatistics (see, e.g., \cite{tolstoy, stoilova, toppan}).\par
%
%
In the present paper, we are interested in an extension of $\Z_2$-graded supersymmetric quantum mechanics (SUSYQM) \cite{cooper, junker} to $\Z_2^2$-graded SUSYQM, as recently proposed by Bruce and Duplij \cite{bruce20a} (see also \cite{toppan, bruce19, aizawa20b}). As shown elsewhere \cite{aizawa20a}, in contrast with standard $\Z_2$-graded SUSYQM, which can be described in terms of ordinary bosons and fermions, $\Z_2^2$-graded mechanics necessitates ordinary bosons, two classes of fermions (commuting among themselves), and exotic bosons that anticommute with the fermions of both classes.\par
%
%
Some years ago, it was shown that standard SUSYQM can be realized in terms of a single bosonic degree of freedom \cite{plyushchay}. Such a minimal bosonization (different from the Nicolai-Witten construction in terms of two bosons \cite{witten, nicolai}) is obtained by using a generalized deformed oscillator algebra (GDOA) \cite{daska, cq96} and imposing a $\Z_2$-grading structure on the deformed bosonic oscillator Fock space, which can be done, for instance, by restricting oneself to a Calogero-Vasiliev algebra \cite{vasiliev}. In such a framework, SUSYQM turns out to be fully reducible, its irreducible components providing two sets of bosonized operators realizing either broken or unbroken SUSYQM. Such an approach in terms of GDOAs was latter on applied \cite{cq02, cq03a, cq03b} to several variants of SUSYQM, namely parasupersymmetric \cite{rubakov, khare93a, beckers90}, orthosupersymmetric \cite{khare93b}, pseudosupersymmetric \cite{beckers95}, and fractional supersymmetric quantum mechanics \cite{durand}.\par
%
%
The purpose of the present work is to extend the approach in terms of GDOAs to $\Z_2^2$-graded SUSYQM.\par
%
%
\section{$\Z_2^2$-graded SUSYQM}

In $\Z_2^2$-graded SUSYQM, we are going to deal with a $\Z_2^2$-graded vector space (over $\C$) that is the direct sum of four homogeneous subspaces, labelled by $\overrightarrow{a} = (a_1,a_2)$, where $a_1$, $a_2 \in \{0,1\}$,
\begin{equation}
  g = g_{(0,0)} \oplus g_{(1,0)} \oplus g_{(0,1)} \oplus g_{(1,1)}.
\end{equation}
Homogeneous elements of $g_{\overrightarrow{a}}$ are denoted by $X_{\overrightarrow{a}}$, $Y_{\overrightarrow{a}}$, \ldots. The vector space $g$ becomes a $\Z_2^2$-graded Lie superalgebra if it admits a bilinear form, denoted by $[[\cdot,\cdot]]$, satisfying the identities
\begin{align}
  & [[X_{\overrightarrow{a}}, Y_{\overrightarrow{b}}]] \in g_{\overrightarrow{a}+\overrightarrow{b}}, \\
  & [[X_{\overrightarrow{a}}, Y_{\overrightarrow{b}}]] = - (-1)^{\overrightarrow{a}.\overrightarrow{b}}
       [[Y_{\overrightarrow{b}}, X_{\overrightarrow{a}}]], \\
  & (-1)^{\overrightarrow{a}.\overrightarrow{c}} [[X_{\overrightarrow{a}}, [[Y_{\overrightarrow{b}},
         Z_{\overrightarrow{c}}]]\, ]] + (-1)^{\overrightarrow{b}.\overrightarrow{a}} [[Y_{\overrightarrow{b}}, 
         [[Z_{\overrightarrow{c}}, X_{\overrightarrow{a}}]]\, ]] \nonumber \\
  & \quad {} + (-1)^{\overrightarrow{c}.\overrightarrow{b}} [[Z_{\overrightarrow{c}}, [[X_{\overrightarrow{a}},
         Y_{\overrightarrow{b}}]]\, ]] = 0, \label{eq:jacobi}
\end{align}   
where 
\begin{equation}
  \overrightarrow{a} + \overrightarrow{b} = (a_1+b_1, a_2+b_2), \qquad \overrightarrow{a}.\overrightarrow{b}
  = \sum_{k=1}^2 a_k b_k
\end{equation}
are calculated modulo 2. Equation (\ref{eq:jacobi}) is called the $\Z_2^2$-graded Jacobi relation. In practice, the general Lie bracket is given by
\begin{equation}
  [[X_{\overrightarrow{a}}, Y_{\overrightarrow{b}}]] = X_{\overrightarrow{a}} Y_{\overrightarrow{b}} -
  (-1)^{\overrightarrow{a}.\overrightarrow{b}} Y_{\overrightarrow{b}} X_{\overrightarrow{a}}
\end{equation}
and therefore coincides with either a commutator or an anticommutator.\par
%
%
In the $\Z_2^2$-graded version of SUSYQM, the $\Z_2^2$-graded superalgebra has four generators $H_{00}$, $Q_{10}$, $Q_{01}$, and $Z_{11}$,\footnote{From now on, for simplicity's sake, we skip all parentheses.} which are Hermitian operators acting in a Hilbert space ${\cal H} = {\cal H}_{00} \oplus {\cal H}_{10} \oplus {\cal H}_{01} \oplus {\cal H}_{11}$ and satisfying the relations
\begin{equation}
  \{Q_{10}, Q_{10}\} = \{Q_{01}, Q_{01}\} = 2H_{00}, \qquad [Q_{10}, Q_{01}] = 2 {\rm i} Z_{11}.
  \label{eq:susy}
\end{equation}
As a consequence of the latter, we also have
\begin{equation}
  [H_{00}, Q_{10}] = [H_{00}, Q_{01}] = [H_{00}, Z_{11}] = 0, \qquad \{Z_{11}, Q_{10}\} = 
  \{Z_{11}, Q_{01}\} = 0.  \label{eq:susy-1}
\end{equation}
The operator $H_{00}$ is the Hamiltonian, $Q_{10}$ and $Q_{01}$ are two supercharges, and $Z_{11}$ is a central element. From now on, we will abbreviate $H_{00}$ and $Z_{11}$ in $H$ and $Z$, respectively.\par
%
%
Instead of $Q_{10}$ and $Q_{01}$, we may consider their linear combinations
\begin{equation}
  Q^{\dagger} = \frac{1}{2}(Q_{10} + {\rm i} Q_{01}), \qquad Q = \frac{1}{2}(Q_{10} - {\rm i}Q_{01}),
\end{equation}
from which we get
\begin{equation}
  Q_{10} = Q^{\dagger} + Q, \qquad Q_{01} = - {\rm i} (Q^{\dagger} - Q).
\end{equation}
On inserting these relations in Eq.~(\ref{eq:susy}), the latter gives rise to the equivalent relations
\begin{align}
  & \{Q^{\dagger}, Q\} = H, \label{eq:susyqm-2a}\\
  & (Q^{\dagger})^2 + Q^2 = 0, \label{eq:susyqm-2b}\\
  & [Q^{\dagger}, Q] = Z.  \label{eq:susyqm-2c}
\end{align} 
Furthermore, Eq.~(\ref{eq:susy-1}) becomes 
\begin{equation}
  [H,Q^{\dagger}] = [H,Q] = [H,Z] = 0, \qquad \{Z,Q^{\dagger}\} = \{Z,Q\} = 0.
\end{equation}
This reformulation of $\Z_2^2$-graded SUSYQM is closer to a usual presentation of standard SUSYQM and will enable us to directly apply some results known for the latter. In the next section, we will proceed to consider the simplest approach of SUSYQM in terms of the Calogero-Vasiliev algebra, leaving another one in terms of a more general GDOA for Sec.~4.\par
%
%
\section{Bosonization in Terms of a Calogero-Vasiliev Algebra}

Let us first recall that the SUSYQM superalgebra, generated by $H$, $Q^{\dagger}$, and $Q$, such that
\begin{equation}
  \{Q^{\dagger}, Q\} = H, \qquad (Q^{\dagger})^2 = Q^2 = 0, \qquad [H, Q^{\dagger}] = [H,Q] = 0,
  \label{eq:susyqm}
\end{equation}
can be realized in terms of the Calogero-Vasiliev algebra, defined by \cite{vasiliev}
\begin{equation}
  [N, a^{\dagger}] = a^{\dagger}, \qquad [N, a] = -a, \qquad [a, a^{\dagger}] = 1 + \kappa T,
\end{equation}
where $N = N^{\dagger}$, $a = (a^{\dagger})^{\dagger}$, $T = e^{{\rm i}\pi N} = (-1)^N$, and $\kappa$ is some real deformation parameter. The operator $T$ is known as the Klein operator. On assuming the existence of a Fock-type vacuum state $|0\rangle$, such that
\begin{equation}
  N |0\rangle = a |0\rangle = 0, \qquad T |0\rangle = |0\rangle,
\end{equation}
the Calogero-Vasiliev algebra admits a Fock space representation, whose carrier space is spanned by the orthonormalized vectors
\begin{equation}
  |n\rangle = \frac{1}{\sqrt{[n]_{\kappa}!}} (a^{\dagger})^n |0\rangle, \qquad \langle n'|n\rangle = \delta_{n',n},
\end{equation}
where $[n]_{\kappa}! = \prod_{i=1}^n [i]_{\kappa}$, $[i]_{\kappa} = i + \frac{\kappa}{2}\bigl(1-(-1)^i\bigr)$. These vectors are such that $N |n\rangle = n |n\rangle$, $a|n\rangle = \sqrt{[n]_{\kappa}} |n-1\rangle$, and $a^{\dagger} |n\rangle = \sqrt{[n+1]_{\kappa}} |n+1\rangle$. One may write
\begin{equation}
  a^{\dagger} a = N + \kappa P_1, \qquad a a^{\dagger} = N + 1 + \kappa P_0, \qquad N = \frac{1}{2} 
  \{a^{\dagger}, a\} - \frac{1}{2}(\kappa + 1),  \label{eq:C-V}
\end{equation}
where $P_0 = \frac{1}{2}(1+T)$ and $P_1 = \frac{1}{2}(1-T)$ are the projectors on even and odd number states, respectively.\par
%
%
The superalgebra (\ref{eq:susyqm}) may be realized in terms of the Calogero-Vasiliev algebra in two different ways \cite{plyushchay}, namely
\begin{equation}
  Q^{\dagger}_0 = a P_0, \qquad Q_0 = a^{\dagger} P_1, \qquad H_0 = \frac{1}{2}\{a^{\dagger}, a\}
  - \frac{1}{2} T [a, a^{\dagger}] = N + \frac{1}{2} - \frac{1}{2} T,  \label{eq:unbroken}
\end{equation}
and
\begin{equation}
  Q^{\dagger}_1 = a P_1, \qquad Q_1 = a^{\dagger} P_0, \qquad H_1 = \frac{1}{2}\{a^{\dagger}, a\}
  + \frac{1}{2} T [a, a^{\dagger}] = N + \kappa + \frac{1}{2} + \frac{1}{2} T.  \label{eq:broken}
\end{equation}
The corresponding (linear) spectrum of $H_0$ and $H_1$ is given by
\begin{equation}
  {\cal E}^{(0)}_n = 2 \left[\frac{n+1}{2}\right]_{\kappa}, \qquad n = 0, 1, 2, \ldots,
\end{equation}
and
\begin{equation}
  {\cal E}^{(1)}_n = 2 \left[\frac{n}{2}\right]_{\kappa} + 1 + \kappa, \qquad n = 0, 1, 2, \ldots,
\end{equation}
respectively. This means that ${\cal E}^{(0)}_0 = 0$, ${\cal E}^{(0)}_{2k+1} = {\cal E}^{(0)}_{2k+2} = 2k+2$, $k=0$, 1, 2, \ldots, whereas ${\cal E}^{(1)}_{2k} = {\cal E}^{(1)}_{2k+1} = 2k+1+\kappa$, $k=0$, 1, 2, \ldots. The former case corresponds to unbroken SUSYQM, while the latter is associated with broken SUSYQM. This has led to the conclusion \cite{plyushchay} that in the framework of Calogero-Vasiliev algebra, SUSYQM is both minimally bosonized and fully reducible.\par
%
%
On comparing Eq.~(\ref{eq:susyqm}) with Eqs.~(\ref{eq:susyqm-2a}) and (\ref{eq:susyqm-2b}), it is obvious that the sets $Q^{\dagger}_0$, $Q_0$, $H_0$ and $Q^{\dagger}_1$, $Q_1$, $H_1$ of Eqs.~(\ref{eq:unbroken}) and (\ref{eq:broken}) also provide a realization of the $\Z_2^2$-graded superalgebra. It only remains to use Eq.~(\ref{eq:susyqm-2c}) to determine the corresponding $Z$ operator. The results read
\begin{equation}
  Z_0 = - T H_0 \qquad \text{and} \qquad Z_1 = T H_1,  \label{eq:Z}
\end{equation}
respectively. As a consequence, in both scenarii, the central element distinguishes between the degenerate eigenstates of $H$ in the Fock space representation, since its eigenvalues are given by
\begin{equation}
  {\cal Z}^{(0)}_0 = 0, \qquad {\cal Z}^{(0)}_{2k+1} = - {\cal Z}^{(0)}_{2k+2} = 2k+2, \qquad k=0, 1, 2, \ldots,
\end{equation}
and
\begin{equation}
  {\cal Z}^{(1)}_{2k} = - {\cal Z}^{(1)}_{2k+1} = 2k + 1 + \kappa, \qquad k=0, 1, 2, \ldots,
\end{equation}
respectively.\par
%
%
\section{Bosonization in Terms of a GDOA}

A more general realization of the SUSYQM superalgebra (\ref{eq:susyqm}) can be obtained by replacing the Calogero-Vasiliev algebra by a GDOA, defined by \cite{daska, cq96}
\begin{equation}
  [N, a^{\dagger}] = a^{\dagger}, \qquad [N, a] = - a, \qquad [a, a^{\dagger}] = G(N),  \label{eq:GDOA}
\end{equation}
where $N = N^{\dagger}$, $a = (a^{\dagger})^{\dagger}$, and $G(N) = [G(N)]^{\dagger}$ is some Hermitian function of $N$. We restrict ourselves to GDOAs possessing a bosonic Fock space representation. In the latter, we may write $a^{\dagger} a = F(N)$, $a a^{\dagger} = F(N+1)$, where the structure function $F(N) = [F(N)]^{\dagger}$ is such that
\begin{equation}
  G(N) = F(N+1) - F(N)
\end{equation}
and is assumed to satisfy the conditions
\begin{equation}
  F(0) = 0, \qquad F(n) > 0 \qquad \text{if $n=1, 2, 3, \ldots.$}
\end{equation}
The carrier space ${\cal F}$ of such a representation can be constructed from a vacuum state $|0\rangle$ (such that $a |0\rangle = N |0\rangle = 0$) by successive applications of $a^{\dagger}$ and its orthonormalized basis states
\begin{equation}
  |n\rangle = \left(\prod_{i=1}^n F(i)\right)^{-1/2} (a^{\dagger})^n |0\rangle, \qquad n=0, 1, 2, \ldots,  
\end{equation}
satisfy the relations $N |n\rangle = n |n\rangle$, $a^{\dagger} |n\rangle = \sqrt{F(n+1)}\, |n+1\rangle$, and $a|n\rangle = \sqrt{F(n)}\, |n-1\rangle$.\par
%
%
A $\Z_2$-grading structure can be imposed on $\cal F$ by introducing a grading operator $T = e^{{\rm i}\pi N} = (-1)^N$ with the two eigenvalues $(-1)^{\mu}$, $\mu = 0$, 1. The two corresponding eigenspaces ${\cal F}_0$ and ${\cal F}_1$, made of even or odd number states, respectively, are such that ${\cal F} = {\cal F}_0 \oplus {\cal F}_1$, the corresponding projectors being $P_0 = \frac{1}{2}(1+T)$ and $P_1 = \frac{1}{2}(1-T)$.\par
%
%
On starting from the general results obtained for order-$p$ parasupersymmetric quantum mechanics in \cite{cq02} and setting $p=1$, which gives back SUSYQM, we arrive at two different realizations of superalgebra (\ref{eq:susyqm}) in terms of the GDOA (\ref{eq:GDOA}), namely
\begin{equation}
  Q^{\dagger}_0 = f(N) a^{\dagger} P_1, \quad Q_0 = f(N+1) a P_0, \quad H_0 = f^2(N) F(N) P_0 +
  f^2(N+1) F(N+1) P_1,  \label{eq:unbroken-bis}
\end{equation}
and
\begin{equation}
  Q^{\dagger}_1 = f(N) a^{\dagger} P_0, \quad Q_1 = f(N+1) a P_1, \quad H_0 = f^2(N) F(N) P_1 +
  f^2(N+1) F(N+1) P_0.  \label{eq:broken-bis}
\end{equation}
Here $f(N)$ is an arbitrary real function of $N$ defined on the set $\{1, 2, 3, \ldots\}$. The eigenvalues of $H_0$ and $H_1$ are given by
\begin{equation}
  {\cal E}^{(0)}_0 = 0, \qquad {\cal E}^{(0)}_{2k+1} = {\cal E}^{(0)}_{2k+2} = f^2(2k+2) F(2k+2), \qquad
  k=0, 1, 2, \ldots,
\end{equation}
and 
\begin{equation}
  {\cal E}^{(1)}_{2k} = {\cal E}^{(1)}_{2k+1} = f^2(2k+1) F(2k+1), \qquad k=0, 1, 2, \ldots,
\end{equation}
respectively. and therefore correspond to either unbroken or broken SUSYQM again. This time, however, the spectra are in general nonlinear.\par
%
%
As already observed in Sec.~3, the operators (\ref{eq:unbroken-bis}) and (\ref{eq:broken-bis}) also provide realizations of the $\Z_2^2$-graded superalgebra  equations (\ref{eq:susyqm-2a}) and (\ref{eq:susyqm-2b}). The calculation of $Z$, given in Eq.~(\ref{eq:susyqm-2c}), now leads to
\begin{equation}
  Z_0 = - f^2(N+1) F(N+1) P_1 + f^2(N) F(N) P_0 \label{eq:Z-unbroken}
\end{equation}
and
\begin{equation}
  Z_1 = - f^2(N+1) F(N+1) P_0 + f^2(N) F(N) P_1.  \label{eq:Z-broken}
\end{equation}
By comparing Eqs.~(\ref{eq:Z-unbroken}) and (\ref{eq:Z-broken}) with (\ref{eq:unbroken-bis} and (\ref{eq:broken-bis}), it is then straightforward to show that
\begin{equation}
  Z_{\mu} = (-1)^{\mu} T H_{\mu}, \qquad \mu = 0, 1.  \label{eq:Z-bis}
\end{equation}
Hence, the eigenvalues of $Z_0$ and $Z_1$ are given by
\begin{equation}
  {\cal Z}^{(0)}_0 = 0, \qquad {\cal Z}^{(0)}_{2k+1} = - {\cal Z}^{(0)}_{2k+2} = - f^2(2k+2) F(2k+2), \qquad 
  k = 0, 1, 2, \ldots,
\end{equation}
and
\begin{equation}
  {\cal Z}^{(1)}_{2k} = - {\cal Z}^{(1)}_{2k+1} = - f^2(2k+1) F(2k+1), \qquad k=0, 1, 2, \ldots,
\end{equation}
respectively. We conclude that the central element distinguishes between degenerate eigenstates of $H$ again.\par
%
%
Whenever the GDOA reduces to the Calogero-Vasiliev algebra, i.e., $G(N) = 1 + \kappa T$, Eq.~(\ref{eq:C-V}) leads to $F(N) = N + \kappa P_1$ and $F(N+1) = N + 1 + \kappa P_0$. On choosing in addition $f(N) = 1$, Eqs.~(\ref{eq:unbroken-bis}), (\ref{eq:broken-bis}), and (\ref{eq:Z-bis}) give back Eqs.~(\ref{eq:unbroken}), (\ref{eq:broken}), and (\ref{eq:Z}) with $Q \leftrightarrow Q^{\dagger}$ and $Z \leftrightarrow -Z$. Such discrepancies are of course irrelevant.\par
%
%
\section{Conclusion}

In this work, we have shown that the minimal bosonization of the $\Z_2$-graded SUSYQM superalgebra in terms of the Calogero-Vasiliev algebra or, more generally, of a GDOA can be extended to the $\Z_2^2$-graded one in a straightforward way. For the latter, the central element $Z$ has the property of distinguishing between degenerate eigenstates of the Hamiltonian.\par
%
%
An interesting open question for future investigation would be the search for a bosonization of the $\Z_2^3$-graded SUSYQM superalgebra in the framework of GDOAs.\par
%
%
\section*{Acknowledgements}

This work was supported by the Fonds de la Recherche Scientifique--FNRS under Grant No.\ 4.45.10.08.\par
%
%

\end{document}